\documentclass{INTERSPEECH2023}

\usepackage{array}
\usepackage{amsfonts}
\usepackage{tikz}
\usepackage{amsmath}
\usepackage{caption}
\usepackage{comment}
\usepackage{color}
\usepackage{booktabs}
\usepackage{xspace}
\usepackage{amssymb}
\usepackage{amsfonts}           
\usepackage{mathrsfs} 
\usepackage{multirow}
\usepackage{booktabs}
\usepackage{graphicx}
\usepackage{enumitem}
\usepackage{mathtools}
\usepackage{cite}
\usepackage{siunitx}
\usepackage{pgfplots}
\usepackage{subcaption}

\interspeechcameraready

\title{Delay-penalized CTC implemented based on Finite State Transducer}

\name{
Zengwei Yao\textsuperscript{*},
Wei Kang\textsuperscript{*},
Fangjun Kuang,
Liyong Guo,\\
Xiaoyu Yang,
Yifan Yang,
Long Lin,
Daniel Povey, \thanks{* stands for equal contribution}}
\address{Xiaomi Corp., Beijing, China}
\email{\{yaozengwei, kangwei1, dpovey\}@xiaomi.com}

\begin{document}

\maketitle

\begin{abstract}

Connectionist Temporal Classification (CTC) suffers from the latency problem when applied to streaming models. We argue that in CTC lattice, the alignments that can access more future context are preferred during training, thereby leading to higher symbol delay. In this work we propose the delay-penalized CTC which is augmented with latency penalty regularization. We devise a flexible and efficient implementation based on the differentiable Finite State Transducer (FST). Specifically, by attaching a binary attribute to CTC topology, we can locate the frames that firstly emit non-blank tokens on the resulting CTC lattice, and add the frame offsets to the log-probabilities. Experimental results demonstrate the effectiveness of our proposed delay-penalized CTC, which is able to balance the delay-accuracy trade-off. Furthermore, combining the delay-penalized transducer enables the CTC model to achieve better performance and lower latency. Our work is open-sourced and publicly available\footnote{https://github.com/k2-fsa/k2}.
\end{abstract}

\vspace{-1mm}
\section{Introduction}

Connectionist Temporal Classification (CTC)~\cite{ctc} is still a competitive choice among various end-to-end models~\cite{transducer, las, deepspeech2, hybrid-ctc-attention, comparison, pruned-rnnt} in Automatic Speech Recognition (ASR) due to its structural simplicity and computational efficiency. CTC aims to maximize the total log-probability over all enumerated alignments between feature sequence and label sequence. However, since CTC equally treats all alignments, it suffers from the symbol delay problem for training streaming ASR models~\cite{bayes-ctc}. It learns to strengthen the higher-score alignments that access more context, which results in high latency when deployed to real applications. 

To address the CTC latency problem, a typical solution is to impose restrictions on the learned alignments by leveraging the ground-truth alignments~\cite{cd-ctc-smbr,fast_and_accurate, StableEmit}. The crucial limitation of this type of methods is that it needs to pre-compute the token-time alignments, which makes the end-to-end training procedure more redundant. Recently, some researches explore reducing the symbol delay without the need of external alignments~\cite{peak-first-ctc, trimtail, bayes-ctc}. 
As a prominent example, Bayes risk CTC~\cite{bayes-ctc} divides all alignments into different groups according to their specific symbol delay, and assigns higher risk values for those lower-delay groups during training. Our method can be explained as a form of Bayes risk CTC, but implemented in a simpler manner based on differentiable Finite State Transducer (FST), instead of modifying the CTC forward-backward algorithm~\cite{ctc} as in Bayes risk CTC~\cite{bayes-ctc}. 

Our work is inspired by delay-penalized transducer~\cite{delay-penalized-rnnt}, which adds a small constant $\lambda$ times the frame offsets to the log-probabilities of emitting symbols, functioning as reducing the averaged symbol delay of transducer lattice. In this paper, we propose a similar method of delay penalization for CTC, which is implemented based on the differentiable FST~\cite{sak2015learning, eesen, nvidia-ctc}. 
Specifically, we first attach a binary attribute to the CTC topology~\cite{nvidia-ctc} indicating whether the arcs are with non-epsilon output labels. The propagated binary attribute on the resulting CTC lattice enables us to easily locate the frames that firstly emit non-blank tokens, so as to add the frame offsets to the log-probabilities. As a result, we can optimize the CTC objective function with latency penalty regularization by maximizing the total score of the modified CTC lattice. Experimental results validate the effectiveness of the proposed delay-penalized CTC implemented based on FST, which provides a way to adjust the trade-off between latency and performance by tuning $\lambda$.

Meanwhile, we also explore leveraging the delay-penalized transducer~\cite{delay-penalized-rnnt} as an auxiliary training task to achieve a better delay-accuracy trade-off for the CTC model. Experimentally, the CTC modeling capacity can be significantly improved with the shared encoder network owing to the superior performance of transducer~\cite{transducer, transformer-transducer, pruned-rnnt}. Compared to the commonly equipped attention decoder~\cite{hybrid-ctc-attention, espnet, wenet}, the essential advantage of the delay-penalized transducer is that it can force the encoder probability distribution to shift to the left, and thereby further reduce the symbol delay of the CTC head. 

Our contributions are as follows:
\begin{itemize}
[leftmargin=*,topsep=1.8pt,parsep=0pt,itemsep=1.8pt,]
\item We propose the delay-penalized CTC, which can be used to train low-latency streaming models without external reference alignments.   
\item A simple and effective implementation based on differentiable FST is designed specifically for the proposed delay-penalized CTC. 
\item Experimentally, combining the delay-penalized transducer in training enables the CTC model to achieve both lower latency and better performance.
\end{itemize}

\vspace{-2mm}
\section{Connectionist Temporal Classification}
\vspace{-1mm}

Given a feature sequence $\mathbf{x} = \{x_t\}_0^{T-1}$ of length $T$, and a label sequence $\mathbf{y} = \{y_u \in \mathcal{V}\}_0^{U-1}$ of length $U$, where $\mathcal{V}$ is the vocabulary, CTC~\cite{ctc} aims to maximize the log-probability $\log P(\mathbf{y}|\mathbf{x})$. By introducing a blank token $\varnothing$, CTC defines the alignments between $\mathbf{x}$ and $\mathbf{y}$: $\mathbf{\pi} = \{\pi_t\}^{T-1}_0$, $\pi_t \in \mathcal{V} \cup \{\varnothing\}$.
An encoder network followed by a linear projection layer and a log-softmax function is utilized to estimate the log-probabilities $\log P (\pi_t)$.

Let $\mathcal{B}(\pi)=\mathbf{y}$ denote a many-to-one map that merges repeated contiguous tokens and removes $\varnothing$. The CTC objective function is to maximize the total log-probability $\mathcal{L}$ over all valid alignments $\pi \in \mathcal{B}^{-1}(\mathbf{y})$:
\begin{equation}
\vspace{-1mm}
\label{eq:ctc_L}
\mathcal{L} = \log \sum_{\pi} \exp(s_{\pi}),
\end{equation}
where $s_{\pi}$ is the log-probability of alignment $\pi$:
\begin{equation}
\label{eq:s_i}
s_{\pi} = \sum_t \log P (\pi_t).
\end{equation}
Typically, $\mathcal{L}$ is efficiently computed employing the forward-backward algorithm~\cite{ctc}. 

Similar to transducer~\cite{delay-penalized-rnnt}, CTC also encounters the latency problem when applied to streaming ASR models~\cite{bayes-ctc}. For example, in streaming Conformer models~\cite{conformer} trained with chunk-wise attention mask, the learned log-probabilities $\log P (\pi_t)$ only access limited future context. On the resulting CTC lattice, the higher-delay alignments that access more context would earn higher scores compared to the lower-delay counterparts, which leads to high recognition latency during decoding.

\vspace{-2mm}
\section{Delay-penalized CTC}

We will first present the formulation of the proposed delay-penalized CTC in section~\ref{ssec:formulation}. Then we will elaborate on the detailed FST-based implementation in section~\ref{ssec:fst-impl}. Finally we will introduce how we utilize the delay-penalized transducer for a better delay-accuracy trade-off in section~\ref{ssec:with_dp_rnnt}.  

\vspace{-1mm}
\subsection{Formulation}
\label{ssec:formulation}

Following the delay-penalized transducer~\cite{delay-penalized-rnnt}, we aim to similarly regularize the CTC objective function $\mathcal{L}$ with an extra term $\mathcal{L}_{\text{delay}}$:
\begin{equation}
\vspace{-0.5mm}
\label{eq:L_aug}
\mathcal{L}_{\text{aug}} = \mathcal{L} + \mathcal{L}_{\text{delay}}.
\end{equation} 
Herein, $\mathcal{L}_{\text{delay}}$ is the averaged delay scores (inversely proportional to symbol delay) of the CTC lattice, scaled by a hyper-parameter $\lambda$:
\begin{equation}
\label{eq:L_delay}
\mathcal{L}_{\text{delay}} = \lambda \sum_{\pi} w_{\pi} \cdot d_{\pi} , 
\end{equation}
where $d_{\pi}$ denotes the delay score of alignment $\pi$, $w_{\pi}$ denotes the normalized alignment weight over the whole lattice: 
\vspace{-0.5mm}
\begin{equation}
\vspace{-1mm}
\label{eq:w_pi}
w_{\pi} = \frac{\exp{(s_{\pi})}}{\sum_{\pi} \exp{(s_{\pi})}}.
\end{equation}
With the augmented objective function $\mathcal{L}_{\text{aug}}$, the CTC model is trained to assign higher log-probabilities on those lower-delay alignments with higher delay scores. 

According to the mathematical proof in~\cite{delay-penalized-rnnt}, for a small $\lambda$, we can approximately compute $\mathcal{L}_{\text{aug}}$ by replacing alignment scores $s_{\pi}$ with $s'_{\pi} = s_{\pi} + \lambda \cdot d_{\pi}$ in the regular CTC function $\mathcal{L}$ in \eqref{eq:ctc_L}:
\begin{equation}
\label{eq:L_aug_approx}
\mathcal{L}_{\text{aug}} \approx \log \sum_{\pi} \exp(s_{\pi} + \lambda \cdot d_{\pi}).
\end{equation}
It can be rearranged as a Bayes risk CTC formula~\cite{bayes-ctc}:
\begin{equation}
\vspace{-1mm}
\label{eq:L_aug_approx_bayes}
\mathcal{L}_{\text{aug}} \approx \log \sum_{\pi} \exp(s_{\pi}) \cdot \exp(\lambda \cdot d_{\pi}).
\end{equation}
where the delay score term $\exp(\lambda \cdot d_{\pi})$ is the Bayes risk function for alignment $\pi$. 

Let $\mathbf{q}=\{q_u\}_0^{U-1}$ denote the frame indexes that firstly emit non-blank tokens $\mathbf{y} = \{y_u\}_0^{U-1}$. The delay score of alignment $\pi$ is defined as:
\begin{equation} 
\label{eq:d_pi}
 d_{\pi} = \sum_u \left(\frac{T-1}{2} - q_u\right),
\end{equation}
where the middle-frame offset $\frac{T-1}{2}$ is used to keep the delay penalty in a proper numerical range. 
From \eqref{eq:s_i}, \eqref{eq:L_aug_approx} and \eqref{eq:d_pi}, before calculating $\mathcal{L}_{\text{aug}}$ with the forward-backward algorithm~\cite{ctc}, we just need to locate those frames $\mathbf{q}$ that firstly emit non-blank tokens and add corresponding frame offsets to the CTC log-probabilities:
\vspace{-2mm}
\begin{equation}
\vspace{-2mm}
\label{eq:penalized_logp}
\log P'(\pi_{q_u}) = \log P(\pi_{q_u}) + \lambda \cdot \left(\frac{T-1}{2} - q_u\right).
\end{equation}
However, to the best of our knowledge, it is a hassle to integrate this change in regular CTC implementations, such as PyTorch-based CTC\footnote{https://github.com/pytorch/pytorch} and warp-ctc\footnote{https://github.com/baidu-research/warp-ctc}. 

\vspace{-1mm}
\subsection{Finite State Transducer-based Implementation}
\label{ssec:fst-impl}

\begin{figure*}[t]
\vspace{-5mm}
     \centering
     \begin{subfigure}[b]{0.45\textwidth}
         \centering
         \includegraphics[width=\textwidth]{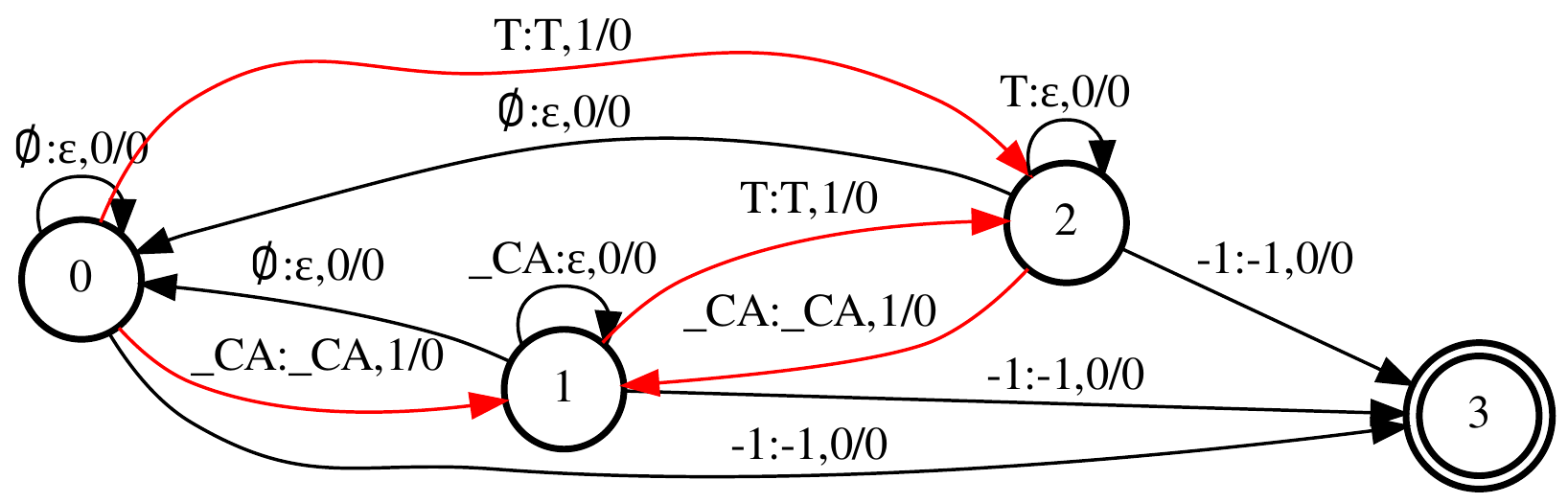}
         \caption{CTC topology $\mathbf{H}$}
     \end{subfigure}
     \begin{subfigure}[b]{0.4\textwidth}
         \centering
         \includegraphics[width=\textwidth]{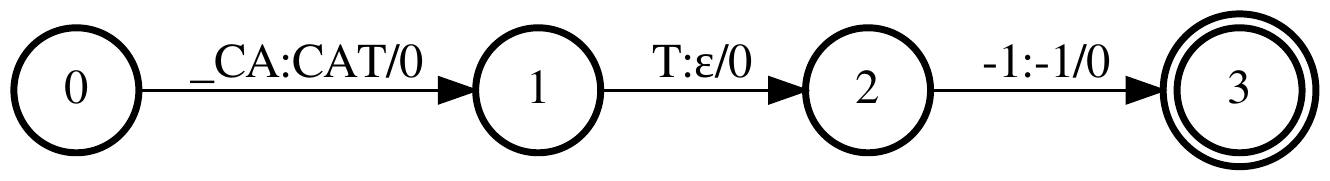}
         \caption{Lexicon $\mathbf{L}$}
     \end{subfigure}
     \bigskip
     \begin{subfigure}[b]{0.6\textwidth}
         \centering
         \includegraphics[width=\textwidth]{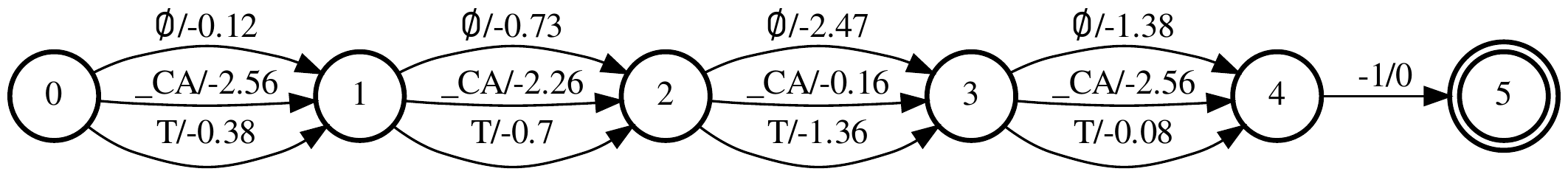}
         \caption{Dense FSA $\mathbf{U}$}
     \end{subfigure}
     \bigskip
     \begin{subfigure}[b]{0.85\textwidth}
         \centering
         \includegraphics[width=\textwidth]{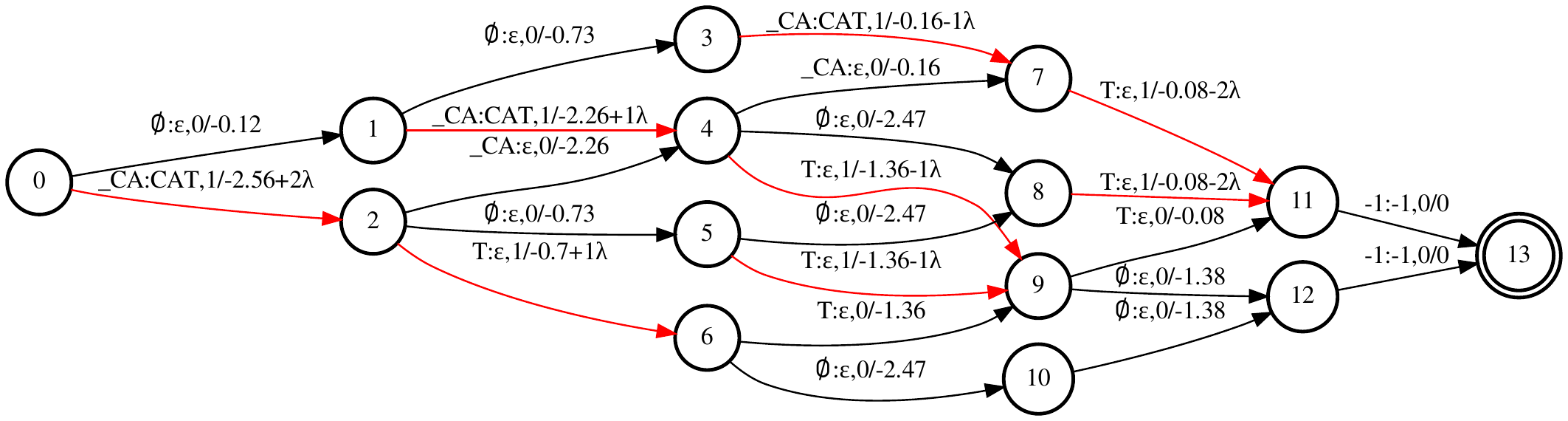}
         \caption{CTC lattice $\mathbf{HLU}$}
     \end{subfigure}
     \vspace{-5mm}
\caption{Delay-penalized CTC implemented based on FST. It visualizes the BPE tokens of "CAT". We attach a binary attribute to $\mathbf{H}$, where $\mathbf{1}$ and $\mathbf{0}$ indicate the arcs with and without non-epsilon ($\epsilon$) output labels, respectively. The arcs with non-epsilon output labels are highlighted in red.
For example, the red arc from state 0 to state 2 on $\mathbf{H}$ with label T:T,$\mathbf{1}$/0, represents input token T and output token T with the binary attribute of $\mathbf{1}$ and score of 0.  On the resulting CTC lattice $\mathbf{HLU}$, we can locate those arcs (in red) that firstly emit non-blank tokens with the attached binary attribute, and add corresponding frame offsets to the log-probabilities. 
}
\label{fig:delay_penalized_HLU}
\vspace{-5mm}
\end{figure*}

We leverage k2 framework\footnote{https://github.com/k2-fsa/k2} to implement the proposed delay-penalized CTC in a simple and efficient way based on the differentiable Finite State Transducer (FST). The FST-based CTC training procedure~\cite{sak2015learning, eesen, nvidia-ctc} typically involves three components:
CTC topology (denoted as $\mathbf{H}$, in Figure~\ref{fig:delay_penalized_HLU}(a)\footnote{Note: In k2 framework, arcs entering the final state always have -1 as label.}), lexicon (denoted as $\mathbf{L}$, in Figure~\ref{fig:delay_penalized_HLU}(b)), and Dense Finite State Acceptor (FSA)\footnote{Search for DenseFsaVec in https://github.com/k2-fsa/k2 for more details.} (denoted as $\mathbf{U}$~\cite{lattice_U}, in Figure~\ref{fig:delay_penalized_HLU}(c)). Specifically, $\mathbf{H}$ serves as the map $\mathcal{B}(\pi)=\mathbf{y}$ that merges repetitions and removes $\varnothing$. 
$\mathbf{L}$ converts sequence of tokens in $\mathcal{V}$ into word sequence. The weights on $\mathbf{U}$ correspond to the acoustic log-probabilities. 
Thus we can construct the CTC lattice (denoted as $\mathbf{HLU}$, in Figure~\ref{fig:delay_penalized_HLU}(d)) which contains all valid alignments $\mathcal{B}^{-1}(\mathbf{y})$ by the following FST operations:
\vspace{-0.5mm}
\begin{equation}
\vspace{-0.5mm}
\label{eq:HLU}
\mathbf{HLU} = \text{intersect} (\text{compose}(\mathbf{H}, \mathbf{L}), \mathbf{U}).
\end{equation}
We can get the CTC loss value and propagate the gradients back to acoustic log-probabilities by computing the total score of $\mathbf{HLU}$ based on dynamic programming algorithm in a differentiable manner.\footnote{Search for Fsa.get\_tot\_scores in https://github.com/k2-fsa/k2 for more details.}


In order to identify the frame indexes that firstly emit non-blank tokens, to which the corresponding penalty scores as in \eqref{eq:penalized_logp} will add, we attach a binary attribute to the arcs in CTC topology $\mathbf{H}$ (in Figure~\ref{fig:delay_penalized_HLU}(a)) indicating the presence of non-epsilon output label. 
Thanks to the k2 mechanism, this attribute will be propagated automatically between the FST operations in \eqref{eq:HLU}.
Hence, on the resulting lattice $\mathbf{HLU}$, we are able to easily locate those arcs where non-blank tokens are firstly emitted with the attached attribute, and add corresponding frame offsets to the log-probabilities (in Figure~\ref{fig:delay_penalized_HLU}(d)). 
By maximizing the total score of the modified $\mathbf{HLU}$, the CTC model is regularized to enhance those lower-delay alignments. 


\vspace{-1mm}
\subsection{Jointly Trained with Delay-penalized Transducer}
\label{ssec:with_dp_rnnt}

In this paper, we also explore training the delay-penalized CTC with a delay-penalized transducer~\cite{delay-penalized-rnnt} as an auxiliary task to achieve higher performance and lower latency for the CTC model. Specifically, the CTC head and the transducer head share a common encoder network but own their individual linear projection layers. As in the hybrid CTC/Attention framework~\cite{hybrid-ctc-attention}, conducting the multi-task learning will strengthen the modeling capacity of the shared encoder. More importantly, with the delay-penalized transducer~\cite{delay-penalized-rnnt}, it is expected to shift the output probability distribution from the shared encoder to the left along the time axis, and thereby encourage the CTC head to emit symbols earlier. 

The joint objective function $\mathcal{L}^{\text{joint}}$ is formulated as:
\begin{equation}
\label{eq:L_joint}
\mathcal{L}^{\text{joint}} = \alpha \cdot \mathcal{L}_{\text{aug}}^{\text{ctc}} + \mathcal{L}_{\text{aug}}^{\text{transducer}}
\end{equation}
where $\mathcal{L}_{\text{aug}}^{\text{ctc}}$ (i.e., $\mathcal{L}_{\text{aug}}$ in \eqref{eq:L_aug}) and $\mathcal{L}_{\text{aug}}^{\text{transducer}}$ are the delay-penalized objective functions for the CTC head and the transducer head, respectively. Note that both heads have their own delay penalty scales $\lambda$. The hyper-parameter $\alpha$ is used to make the loss values of two heads numerically close.

\vspace{-1mm}
\section{Experiments}

\begin{figure*}[t]
\vspace{-3mm}
\centering
\includegraphics[width=16cm]{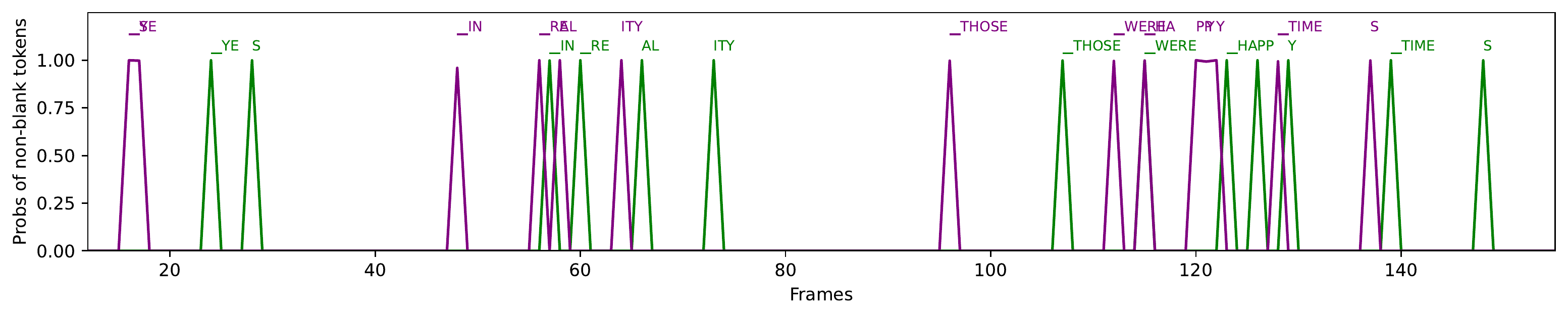}
\caption{Visualization of CTC output probability distributions of non-blank tokens from the streaming models with (in purple) and without (in green) delay penalization, respectively. }
\label{fig:ctc_peak}
\vspace{-3mm}
\end{figure*}

\subsection{Experimental Setup}

\noindent\textbf{Dataset and Evaluation metrics.} We conduct ASR experiments on LibriSpeech dataset~\cite{librispeech}, which contains 1000 hours of English speech with sampling rate of 16 kHz. We evaluate model performance and latency on subsets \textit{test-clean} and \textit{test-other}. Our reference alignments are generated using the torchaudio toolkit~\cite{torchaudio}. Word Error Rate (WER) is used to evaluate the recognition performance. Lower value of WER indicates better recognition result. Two metrics are employed to measure the CTC latency, namely Mean Start Delay (MSD) and Mean End Delay (MED). Specifically, for all correctly recognized words, MSD and MED calculate the averaged start time difference and the averaged end time difference between the prediction and the reference alignment, respectively. Higher value of MSD or MED indicates higher emission latency. CTC greedy search is adopted as the decoding method for a better measure of model performance and symbol delay. 

\noindent\textbf{Implementation Details.} Our data preparation is performed using the Lhotse~\cite{lhotse} framework. Speed perturbation~\cite{speed_perturb} with factors of 0.9 and 1.1 is applied to augment the training data. SpecAugment~\cite{SpecAugment} and MUSAN~\cite{musan}-based noise augmentation are utilized to improve model robustness in training. The input features are 80-channel Fbank extracted on 25 ms windows shifted by 10 ms. The classification units are 500-classes Byte Pair Encoding (BPE)~\cite{bpe} word pieces. We use a 12-layer Conformer~\cite{conformer} with 77 M parameters as the encoder. The input features are subsampled with a factor of 4 by a convolutional layer. The embedding dimension is set to 512. The kernel size of the convolution layers is set to 31. The feedforward dimension is set to 2048. To train the streaming models, the attention layers are trained with block-triangular mask to limit the future context. During streaming inference we use a chunk size of 320 ms. While training with the auxiliary transducer head, we adopt the pruned transducer~\cite{pruned-rnnt} implementation for high efficiency and low memory consumption, in which the stateless decoder~\cite{rnntstateless} is a 1-D convolution layer with a kernel size of 2. The hyper-parameter $\alpha$ in \eqref{eq:L_joint} is set to 0.2. All models are trained for 35 epochs with automatic mixed precision. 

\vspace{-2mm}
\subsection{Results of delay-penalized CTC}
\vspace{-1mm}

\begin{table}[h]
  \centering
  \caption{Results of CTC models in non-streaming and streaming modes.}
  \vspace{-2mm}
  \label{tab:lambda}
  \setlength{\tabcolsep}{2.2pt} 
  \renewcommand{\arraystretch}{1.2}
  \resizebox{1\linewidth}{!}{
  \begin{tabular}
  {l|ccc|ccc}
  \toprule
  \multirow{3}{*}{Method} & \multicolumn{3}{c|}{\textit{test-clean}} & \multicolumn{3}{c}{\textit{test-other}} \\
  & WER & MSD & MED & WER & MSD & MED \\
  & (\%) & (ms) & (ms) & (\%) & (ms) & (ms) \\
  \midrule
  Non-streaming & 3.24 & -28 & -100 & 7.89 & -28 & -100 \\
  Streaming & 4.56 & 273 & 189 & 12.21 & 275 & 192 \\
  \hspace{0.8mm} + Reference Alignment & 5.45 & 117 & 33 & 13.26 & 117 & 34 \\
  \hspace{0.8mm} + Delay penalty, $\lambda=0.005$ & 5.00 & 179 & 97 & 12.86 & 186 & 105 \\
  \hspace{0.8mm} + Delay penalty, $\lambda=0.010$ & 5.32 & 108 & 33 & 13.82 & 121 & 44 \\
  \hspace{0.8mm} + Delay penalty, $\lambda=0.015$ & 5.58 & 42 & -36 & 14.04 & 63 & -17 \\
  \hspace{0.8mm} + Delay penalty, $\lambda=0.020$ & 5.96 & 3 & -75 & 14.81 & 23 & -56 \\
  \hspace{0.8mm} + Delay penalty, $\lambda=0.025$ & 6.44 & -18 & -101 & 15.53 & 2 & -81 \\
  \bottomrule
  \end{tabular} }
\vspace{-4mm}
\end{table}

We first perform experiments to validate the effectiveness of the proposed method of delay penalization on CTC implemented based on FST. Table~\ref{tab:lambda} presents the results of CTC models in non-streaming and streaming modes respectively. Compared to the non-streaming model, the streaming model without delay penalty obtains significantly higher symbol delay, since CTC learns to augment the higher-delay alignments that access more future context during training. We also conduct an experiment to leverage the pre-computed reference token-time alignments for comparison, which performs the frame-wise BPE token classification with a loss scale of 0.75. The result indicates that the streaming model learns to predict non-blank tokens earlier with the frame-wise supervision by the reference alignments. For the streaming models with delay penalty, increasing $\lambda$ consistently leads to lower MSD and MED, and higher WER. It manifests that the proposed method can effectively balance the trade-off between symbol delay and accuracy by tuning $\lambda$. Note that the delay-penalized model with $\lambda=0.010$ achieves similar symbol delay compared to the model using reference alignments, which demonstrates the effectiveness of the proposed regularization method.

Figure~\ref{fig:ctc_peak} visualizes the CTC output probability distributions of non-blank tokens from the streaming models with ($\lambda=0.025$) and without delay penalization, respectively. The CTC peaks shifting to the left along the time axis indicates that the delay penalty regularization enables the CTC model to learn to emit non-blank tokens earlier and thereby achieve lower latency. 

\vspace{-2mm}
\subsection{Results of combining delay-penalized transducer}

Next we conduct experiments to investigate the effect of leveraging the delay-penalized transducer~\cite{delay-penalized-rnnt} as an auxiliary task for the streaming CTC model. We use the commonly adopted hybrid CTC/Attention framework~\cite{hybrid-ctc-attention} for comparison. Specifically, we build a 6-layer transformer-based decoder with 25.7 M model parameters, which serves as a jointly learned language model. The total number of model parameters in the pruned transducer head~\cite{delay-penalized-rnnt} is only 1.0 M. Table~\ref{tab:diff_method} presents the results in terms of symbol delay and recognition accuracy of the CTC models trained with different auxiliary tasks. Training with attention decoder and transducer both improve the CTC recognition performance owing to their modeling ability in learning the dependencies of the output tokens. Attention decoder does not affect the CTC latency, due to its non-streaming modeling mechanism that accesses full encoder context~\cite{hybrid-ctc-attention}. Training with the transducer without delay penalty leads to a higher symbol delay, since the transducer also favors the higher-delay alignments that access more future context~\cite{delay-penalized-rnnt} and thereby affects the probability distribution from encoder. After applying delay penalty on the transducer~\cite{delay-penalized-rnnt}, the CTC models achieve lower latency compared to the baseline. It demonstrates the advantage of combining the delay-penalized transducer over the attention decoder that it can force the encoder probability distribution to shift to the left along the time axis. 

Table~\ref{tab:ctc_rnnt} presents the results of delay-penalized CTC models trained with and without the delay-penalized transducer~\cite{delay-penalized-rnnt} using different $\lambda$, respectively. Combining the delay-penalized transducer consistently improves the recognition performance and reduces the symbol delay for CTC models, which manifests the effectiveness of the proposed multi-task framework. 


\vspace{-2mm}
\begin{table}[h]
  \centering
  \caption{Results of streaming CTC models trained with different auxiliary tasks.}
  \label{tab:diff_method}
  \setlength{\tabcolsep}{3pt} 
  \renewcommand{\arraystretch}{1.2}
  \resizebox{1\linewidth}{!}{
  \begin{tabular}
  {l|ccc|ccc}
  \toprule
  \multirow{3}{*}{Method} & \multicolumn{3}{c|}{\textit{test-clean}} & \multicolumn{3}{c}{\textit{test-other}} \\
  & WER & MSD & MED & WER & MSD & MED \\
  & (\%) & (ms) & (ms) & (\%) & (ms) & (ms) \\
  \midrule
  Baseline & 4.56 & 273 & 189 & 12.21 & 275 & 192 \\
  \hspace{0.8mm} + Attention & 4.15 & 270 & 188 & 10.89 & 273 & 191 \\
  \hspace{0.8mm} + Transducer & 4.00 & 342 & 271 & 10.42 & 345 & 274 \\
  \hspace{0.8mm} + Transducer, $\lambda=0.0050$ & 4.19 & 151 & 72 & 10.95 & 175 & 97 \\
  \hspace{0.8mm} + Transducer, $\lambda=0.0075$ & 4.42 & 135 & 49 & 11.21 & 155 & 72 \\
  \hspace{0.8mm} + Transducer, $\lambda=0.0100$ & 4.51 & 122 & 37 & 11.40 & 141 & 58 \\
  \bottomrule
  \end{tabular} }
  \vspace{-3mm}
\end{table}
\vspace{-3mm}

\begin{table}[h]
  \centering
  \caption{Results of delay-penalized CTC models trained w/o the delay-penalized transducer respectively.}
  \label{tab:ctc_rnnt}
  \setlength{\tabcolsep}{3pt} 
  \renewcommand{\arraystretch}{1.1}
  \resizebox{1\linewidth}{!}{
  \begin{tabular}
  {c|c|ccc|ccc}
  \toprule
  \multirow{3}{*}{$\lambda$ in CTC} & \multirow{3}{*}{$\lambda$ in Transducer} & \multicolumn{3}{c|}{\textit{test-clean}} & \multicolumn{3}{c}{\textit{test-other}} \\
  & & WER & MSD & MED & WER & MSD & MED \\
  & & (\%) & (ms) & (ms) & (\%) & (ms) & (ms) \\
  \midrule
  0.010 & - & 5.32 & 108 & 33 & 13.82 & 121 & 44 \\
  0.010 & 0.0050 & \textbf{4.74} & \textbf{77} & \textbf{-3} & \textbf{11.89} & \textbf{101} & \textbf{22} \\
  \midrule
  0.015 & - & 5.58 & 42 & -36 & 14.04 & 63 & -17 \\
  0.015 & 0.0075 & \textbf{4.91} & \textbf{15} & \textbf{-58} & \textbf{12.05} & \textbf{48} & \textbf{-26} \\
  \midrule
  0.020 & - & 5.96 & 3 & -75 & 14.81 & 23 & -56 \\
  0.020 & 0.0100 & \textbf{5.3} & \textbf{-23} & \textbf{-99} & \textbf{12.76} & \textbf{8} & \textbf{-67} \\
  \bottomrule
  \end{tabular} }
  \vspace{-3mm}
\end{table}
\vspace{-3mm}

\section{Conclusion}
In this work, we propose the delay-penalized CTC, which is implemented based on the differentiable FST. Specifically, for the CTC lattice modeled by FST, we locate the frames that firstly emit non-blank tokens and add the corresponding frame offsets to the CTC log-probabilities. 
Experimental results demonstrate that it can effectively balance the trade-off between symbol delay and recognition performance. Furthermore, leveraging a delay-penalized transducer as the auxiliary task enables the CTC model to achieve a better delay-accuracy trade-off. 

\bibliographystyle{IEEEtran}
\bibliography{mybib}

\end{document}